# DESIGN OF A CAPACITOR-LESS LOW-DROPOUT VOLTAGE REGULATOR

X. R. LI, D.B. PEI, Q. LIU and R.SHEN

*Institute of Electronic CAD, Xidian University,*

*No. 2, South Tai Bai Road,*

*Xi'an 710071, P R China*

lixianrui4213@126.com

A solution to the stability of capacitor-less low-dropout regulators with a 4pF Miller capacitor in Multi-level current amplifier is proposed. With the Miller compensation, a more than 50°phase margin is guaranteed in full load. An extra fast transient circuit is adopted to reduce stable time and peak voltage. When the load changes from light to heavy, the peak voltage is 40mV and chip quiescent current is only 45uA.

## 1. Instruction

Multiple Low-Dropout-Regulators (LDO) are often used for power management in multi-core designs or System-on-Chip (SoC) applications. As the number of LDOs used in such systems may be dozens, the number of on- and off-chip component required by every single LDO directly affects the total chip area, pins of package, and thus cost of the system. However, it is well known that LDO suffers intrinsic stability problem and requires a relatively large capacitor for frequency compensation [1-5]. This compensation capacitor, whether on or off-chip, makes full integration of LDO difficult if not impossible. In response to this challenge, many advanced LDO structures have been proposed [6-8] recently. While being capable to demonstrate outstanding performance, some of them still require off-chip capacitor or more than one on-chip capacitor to stabilize the system, thus is less attractive for SoC application. In view of this, this paper proposes a new capacitor-less LDO structure that favors SoC designs[9-11].

In this paper, a new topology which is based on Multi-level current amplifier combined with a transient response enhancement network is proposed. For a full load range from 1mA to 200mA, it can achieve not only high stability but also fast transient response. Compared with 12pF and 7pF reported in [6] and [9], the total value of required on-chip capacitors is only 4pF, which occupies less silicon area and makes it attractive for full on chip power management.

## 2. Conventional LDO and its Limitation in SoC

The schematic diagram of the conventional LDO is shown in Fig.1. A conventional LDO includes a pass transistor (M1), error amplifier (EA) and feedback network (R1, R2). Once the output voltage changes, the resistors feedback the changes to EA and then adjust the current of pass transistor to make the output voltage stable.

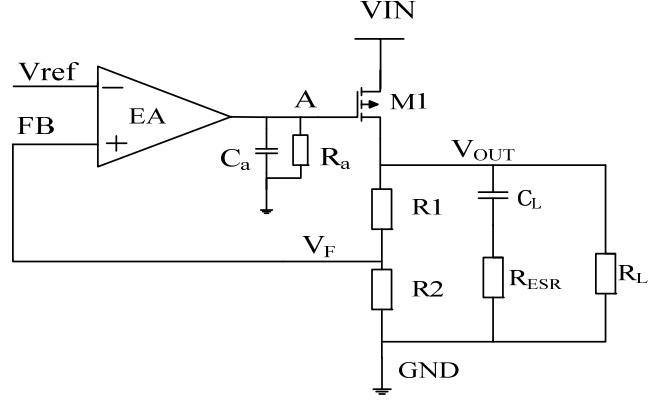

Fig. 1. Structure of conventional LDO.

As Fig. 1 shows, the convention LDO has two poles. The gate capacitance of the power PMOS will contribute a parasitic pole located at low frequencies. Furthermore, because the external load capacitor is very large as well, it will contribute another dominant pole.

$$p_1 = \frac{1}{R_a C_a} \tag{1}$$

$$p_2 = \frac{1}{R_o C_L} \tag{2}$$

Where, Ra is the output impedance of error amplifier, Ca is the parasitic capacitance on gate of pass transistor, Ro is the impedance of output, $C_L$ is the output capacitor. On the other hand, with the series connection of the load capacitor and its equivalent series resistance (ESR), it generates a zero.

$$Z = \frac{1}{R_{ESR} C_L} \tag{3}$$

Where, $R_{ESR}$ is equivalent series resistance of capacitor $C_L$. The conventional method is making use of the equivalent series resistance of the output capacitor to produce a constant zero for compensating the phase shift introduced by the non-dominant pole. However, if the loop gain is too high, and $P_1$ locates before the unity-gain frequency, a bulky tantalum capacitor or a small ceramic capacitor in series with external resistor is required to retain stability for the LDO regulator. This compensation method usually has a narrow bandwidth. Additionally, it may be unstable when load varies largely because of the frequency shift of output pole. For example, the frequency of the output pole will increase about 100 times when the load current varies from 1mA to 100mA. And it will be more serious when the LDO features ultra low power because of the internal high-impedance nodes. Hence, conventional ESR compensation has limited applications for LDO on SOC.

3．The LDO structure for SoC

The small signal model of proposed LDO is shown in Fig. 2. In this figure, the LDO uses the multi-stage current amplifier and a small capacitor compensation Cm inside the system. That makes the system stable. $C_F$ is the output parasitic capacitance, generally is 10 pF to 100 pF.

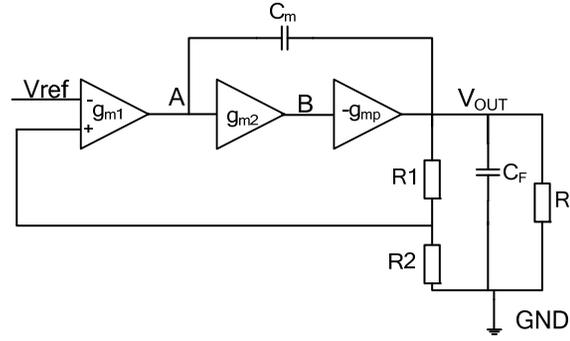

Fig. 2. The structure of proposed LDO.

The overall transfer function is given as

$$H(s) \approx \frac{A_{dc}(1 - \frac{C_m}{g_{m2}g_{mp}R_{o2}}s - \frac{C_m C_p}{g_{m2}g_{mp}}s^2)}{(1 + C_m R_{o1} g_{m2} g_{mp} R_{o2} R_o s)(1 + \frac{C_F}{g_{m2}g_{mp}R_{o2}}s + \frac{C_F C_p}{g_{m2}g_{mp}}s^2)} \qquad (4)$$

Where $gm_1$ and $gm_2$ are the transconductance of the op-amp, and , gmp is the transconductace of the pass transistor, While $Ro_1$, $Ro_2$, Ro, Cm, and Cp are the output resistance of the first, second op-amp, pass transistor, miller compensation capacitor, and the gate parasitic capacitor of the pass transistor, respectively. Adc is the dc loop of the regulator, which can be approximately to

$$A_{dc} = g_{m1} g_{m2} g_{mp} R_{o1} R_{o2} R_o \qquad (5)$$

The poles and zero of (4) can be described as fellow:

$$p_1 = \frac{1}{C_m R_{o1} g_{m2} g_{mp} R_{o2} R_o} \qquad (6)$$

$$p_{2,3} = \frac{-C_F \pm \sqrt{C_F^2 - 4 C_F C_p R_{o2}^2 g_{m2} g_{mp}}}{2 C_F C_p R_{o2}} \qquad (7)$$

$$z_{1,2} = \frac{-C_m \pm \sqrt{C_m^2 + 4 C_m C_p R_{o2}^2 g_{m2} g_{mp}}}{2 C_m C_p R_{o2}} \qquad (8)$$

From the (6) to (8), these make the pole P1 low enough such that become the main pole and the other two poles are located after UGF. Right half plane zero Z1 and left half plane zero Z2 are located in high frequency, so the system only have one pole within the unit gain bandwidth and system is stable.

The corresponding schematic diagram of Fig.2 is illustrated in Fig.3 . Q1 and R1 provide bias current to modules. Q2-Q6 and resistance R2-R7 form the band-gap reference circuit. The first op-amp is error amplifier and band-gap, the second op-amp consists of Q10 to Q13, the third op-amp is the NPN pass transistor. Known from the analysis above, through the compensation of Cm system only have one pole within the unit gain bandwidth, and system is stable.

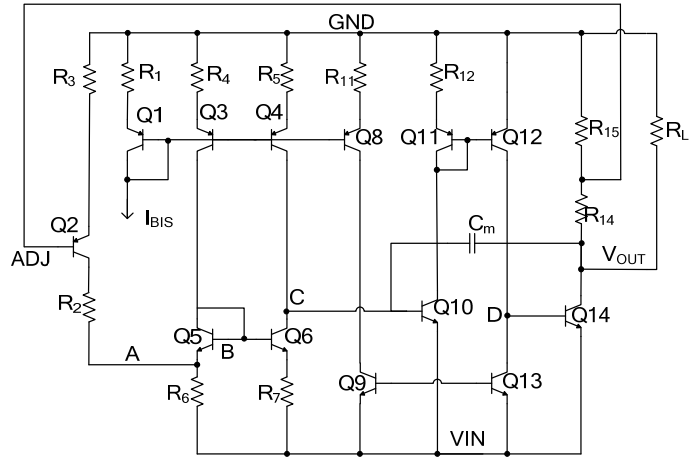

Fig. 3. The proposed schematic of LDO.

**4. The transient enhancement circuit**

By using the internal compensation capacitor can make the system stable, but when the load current change, there is no outside large capacitance to charge and discharge at the moment. Which makes the transient response worse. Therefore, in order to improve the transient performance of the circuit, this article increases the transient enhancement circuit in the system. The schematic is shown in Fig. 4.

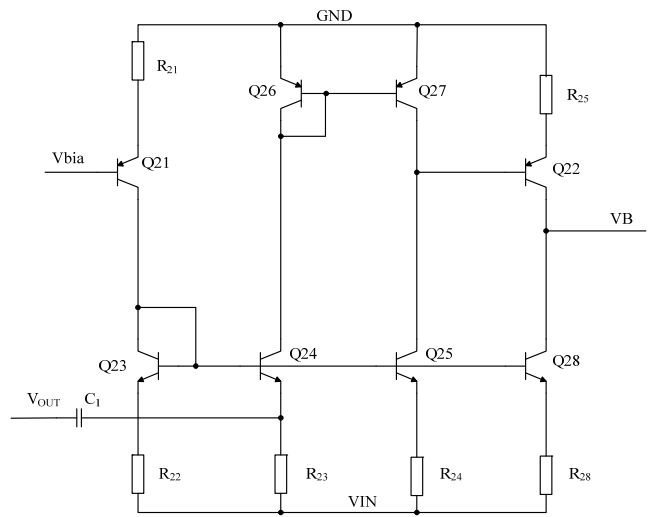

Fig. 4. The circuit of transient enhancement.

In the Fig. 4, $V_{OUT}$ is the output of the LDO. The small of Capacitor C1 is mainly couple the change of $V_{OUT}$ into the emitter of Q24. Vbia is bias voltage from bias module, and makes Q21 on to provide bias current. Transistors Q23- Q25 and R22-R24, R28 compose current mirror. VB is the output terminal of transient response circuit, and connected to the base of NPN pass transistor. This circuit is equivalent to increase a path from the output $V_{OUT}$ to the base of NPN pass transistor. When the output voltage is constant, the capacitor C1 makes the transient enhancement module off. Once the output voltage $V_{OUT}$ changes, capacitor C1 would response the change to transient enhancement circuit immediate, and quickly built a path from the output VOUT to the base of NPN pass transistor directly rather than through a slower EA loop. Due to the design requirements for small power consumption, we can increase the resistance to decrease the static current of transient enhancement module, thereby reduce power consumption.

## 5．The simulation results

The structure of proposed LDO is simulated based on 20V Bipolar process using Cadence software. The Fig. 5(a) is the gain and phase curve in 100mA load current. As Fig. 5(a) shown, the loop gain is 58dB and phase margin is 64° though the Miller compensation. The Fig. 5(b) is the static current curve in 100mA load current at different input voltages. As Fig. 5(b) shown, when the input voltage change, the static current of the circuit is about 45uA, and the power consumption is low.

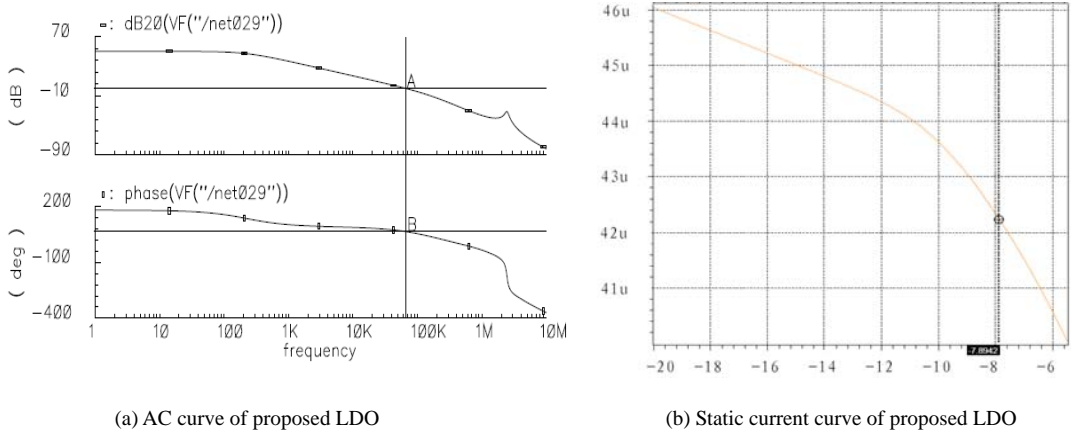

(a) AC curve of proposed LDO            (b) Static current curve of proposed LDO

Fig. 5. The curves of proposed LDO.

Fig. 6 is the load transient response curve at VIN is -15V when load curent changes from 1mA to 100mA. After increasing the transient enhancement circuit, voltage peak is 40mV, the recovery time is 30 us. The proposed LDO had passed the simulation and the chip micrograph is shown in Fig. 7.

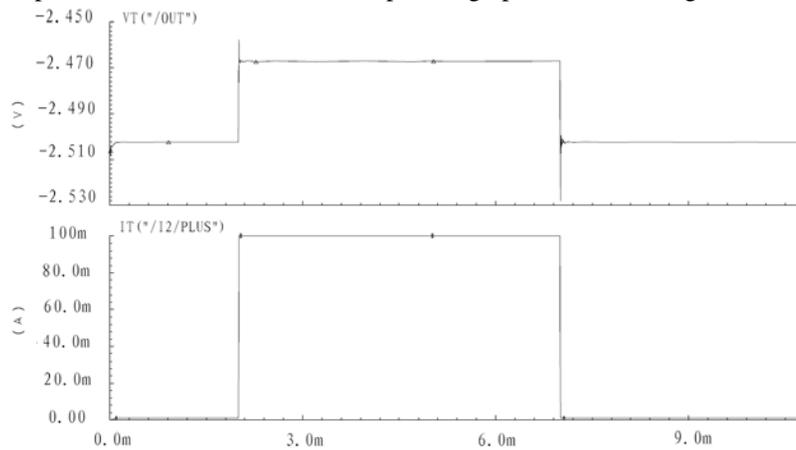

Fig. 6. The load transient response

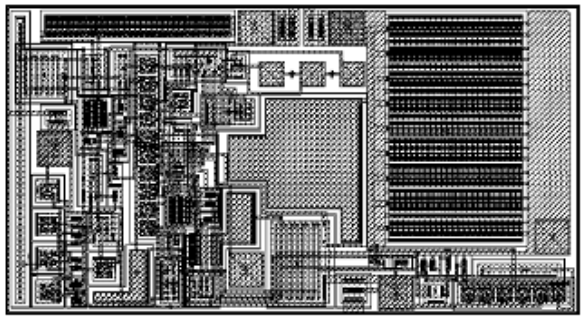

Fig. 7. Chip Micrograph of the proposed LDO.

## 6. Conclusion

This paper proposes a capacitor-less LDO which can apply to in SOC or multi-core design. The proposed capacitor-free LDO makes the system stable in full load with 4pF Miller capacitor. The transient enhancement module makes the peak voltage 40mV when the load current changes from 1mA to 100mA in 10us. By using NPN as pass transistor static current is about 45uA in the input voltage range which realized low power consumption. The proposed LDO had passed the simulation and is tipping out.

**Acknowledgments**

This work was partly supported by the National Natural Science Foundation of China (No.F040202), and by Fundamental Research Funds for the Central Universities of China (No.JB140210).